

\magnification=\magstep1


\vsize=24.0truecm
\hsize=15.5truecm


\vsize=22.5 truecm
\hsize=16.5 truecm


\baselineskip=12pt
\tolerance=5000
\parskip=7pt

\font\grand=cmbx10 at 14.4truept
\font \ninebf                 = cmbx9
\font \ninerm                 = cmr9

\rightline{UdeM-LPN-TH-77/91\break}
\rightline{DIAS-STP-91-42a\break}

\vskip 1.2truecm

\baselineskip=16pt

\centerline{{\grand
Rational vs Polynomial Character of W$_n^l$-Algebras} \footnote{*}
{Revised version of the preprint UdeM-LPN-TH-77/91, DIAS-STP-91-42,
entitled ''Polynomial and Primary Field Character of
W$_n^l$-Algebras''}
\break}

\vskip 2.0truecm

\centerline{L. Feh\'er}
\medskip
\centerline{\it Laboratoire de Physique Nucl\'eaire}
\centerline{\it Universit\'e de Montr\'eal}
\centerline{\it Montr\'eal, Canada H3C 3J7}

\vskip 1truecm

\centerline{L. O'Raifeartaigh, P. Ruelle and I. Tsutsui}
\medskip
\centerline{\it Dublin Institute for Advanced Studies}
\centerline{\it 10 Burlington Road, Dublin 4, Ireland}

\vskip 3.5truecm

\centerline{\bf Abstract}
\medskip
The constraints proposed recently by Bershadsky to produce $W^l_n$
algebras are a mixture of first and second class constraints
and are degenerate. We show that they admit a first-class subsystem
from which they can be recovered by gauge-fixing, and that the
non-degenerate constraints can be handled by previous methods. The
degenerate constraints present a new situation in which the natural
primary field basis for the gauge-invariants is rational rather than
polynomial. We give an algorithm for constructing the rational basis
and converting the base elements to polynomials.

\vfill\eject

\def\d{\delta}
\def\D{\Delta}

\def\G{\Gamma}
\def\s{\sigma}
\def\ss{\otimes}
\def\h{\times}
\def\SS{\Sigma}
\def\S{{\cal D}_0}
\def\ad{{\rm ad}}

\def\\{{\hfil\break }}

\def\rr{\rangle}
\def\ll{\langle}
\def\half{{1 \over 2}}

\def\j{j^{\rm hw}_{{\cal D}_0}}

In recent years it has been found that the W-algebras of
Zamolodchikov (polynomial extensions of the Virasoro
algebra by
primary fields) occur naturally in the context of
linearly constrained Kac-Moody (KM) theory [1-3], and are
the canonical symmetry
algebras of the associated constrained dynamical systems [3-5].
Most results
to date have been obtained for the cases in
which the constraints are
first class and non-degenerate, where non-degenerate means
that no element of the associated gauge-algebra
commutes with the constant (non-zero)
component of the constrained current [3]. However, a
system of constraints proposed recently by
Bershadsky [6], who developed the idea of Polyakov [7] for
the $sl(n,R)$ KM algebra, satisfies neither of the above two
conditions and thus raises the question as to whether the algebras
associated with this new system of constraints, called
$W^l_n$-algebras, are
of the Zamolodchikov kind.
This has already been shown
to be the case for
the simplest example $W_3^2$ [7] and further studied for $W_4^2$
[8] and $W_4^3$ [9] but, as far as we
know, there are no results for the general case. The purpose of
the present note is to provide some general results, as follows:

\noindent
(i) The mixed set of first- and second-class constraints admits a
first-class
subset $\Gamma$ from which they can be recovered by gauge-fixing.

\noindent
(ii) The first-class constraint algebra $\Gamma$ differs from those
previously encountered in that it has a degenerate subalgebra,
i.e. a subalgebra ${\cal D}_0$ such that $[M,{\cal D}_0]=0$, where
$M$
is the constant component of the current. Only for the algebras
$W^2_n$ with odd  $n$ is ${\cal D}_0=0$. In all other cases
$\Gamma$
is a semi-direct sum of the form $\Gamma = {\cal D}_0 \wedge
\tilde\Gamma$.

\noindent
(iii) The system has an $sl(2,R)$ symmetry similar to that
encountered previously. This symmetry allows one to handle the
non-degenerate part
$\tilde \Gamma$ of the algebra by the techniques developed in [1]
and [3] and produces a complete set of primary-field
$\tilde \Gamma$-invariant polynomials $j^{\rm
hw}(j_{\tilde\Gamma})$ in the $\tilde\Gamma$-constrained currents
$j_{\tilde\Gamma}$,
which are highest weights with respect to
$sl(2,R)$. The Poisson bracket algebra of these $j^{\rm hw}$ is a
Zamolodchikov algebra, which is the generalization of those found
for
the $W^2_4$ and $W^3_4$ in [8,9] and is a special case of the
${\cal W}^{\cal G}_{\cal S}$ algebras
considered in [3].

\noindent
(iv) The full set of constraints in [6] include, however, those
corresponding to ${\cal D}_0$. Their inclusion amounts to
eliminating some of the highest weight fields $j^{\rm hw}$,
leaving a subset
$j^{\rm hw}_{{\cal D}_0}$, say.  But,
because of their degeneracy the
${\cal D}_0$-constraints
introduce a new feature, namely
that the natural basis for the
gauge-invariants  is a set of {\it rational}
dim$W^l_n$ functions
$j^{\rm red}(j^{\rm hw}_{{\cal D}_0})$.
This means that the natural Poisson-bracket
algebra for these invariants is a rational, rather than
a polynomial,
extension of the Virasoro algebra and thus is not a Zamolodchikov
algebra in the original sense of the word.

\noindent
(v) Because the gauge-subalgebra ${\cal D}_0$ is scalar with
respect to
$sl(2,R)$ it is possible to convert the rational base-functions
$j^{\rm red}(\j)$ into polynomials $p^{\rm red}(\j)$. But
(as we show by examples) there exist gauge-invariant polynomials
of the $\j$
that are only rational, but not polynomial, functions of the
$p^{\rm red}(\j)$.
Thus the basis remains rational and there is no guarantee that
the Poisson-bracket algebra of the $p^{\rm red}(\j)$ closes
polynomially.

We first recall the general structure of linear constraints [3].
For this it is convenient to write the KM algebra in the form
$$
\{\ll a,J(x) \rr,\ll b,J(y)\rr \}=\ll [a,b],J(y) \rr \, \d(x-y) +
\kappa \ll a,b \rr \,\d'(x-y), \eqno(1)
$$
where $a$ and $b$ are elements of the underlying
finite dimensional simple Lie algebra
${\cal G}$, and $\kappa$, up to a normalization factor,
is the KM level.
(For notational simplicity, we henceforth set $\kappa = 1$ except
in the formulae of Virasoro centre given in the end.)
Letting $\Gamma$ denote
any subalgebra of ${\cal G}$, and $M$ any element of ${\cal
G}$,
the linear constraints bring the current into
the following
form:
$$
J_\G(x)= M + j_\G(x)\ ,
\qquad {\rm with}\qquad j_\G(x) \in \Gamma^\perp\ .
\eqno(2)
$$
A sufficient condition for the
constraints (2) to be conformally invariant is
the existence of an
element $H$ in the Lie algebra ${\cal G}$ such that
$$
[H,M]=-M, \qquad \ll H,\gamma \rr =0 \quad \hbox{and}
\quad [H,\gamma] \in
\Gamma, \quad \forall \gamma \in \Gamma.
\eqno(3)
$$
Indeed,  if  there exists such an $H$, one can verify from (1)
that the following modified Virasoro density
$$
L_H(x)=L_{\rm KM}(x)-\ll H, J'(x)\rr,
\qquad {\rm where}\qquad
L_{\rm KM}(x)={1 \over 2} \ll J(x), J(x) \rr,
\eqno(4)
$$
weakly commutes with the constraints in (2).
The equation $[H,M]=-M$ implies that $M$
is nilpotent, and every nilpotent element of
a real simple Lie
algebra has an $sl(2,R)$ subalgebra,
$\{M_-,M_0,M_+\}$
say, associated with
$M \equiv M_-$. It turns out to be very convenient to use
this $sl(2,R)$ algebra, in which the $M_0$ element can play
the role
of $H$, to describe the constrained system.

If we demand that the constraints in (2) be first class,
they must satisfy the following two conditions
(in addition to $\Gamma$ being a subalgebra) [3]:
$$
\ll \gamma_i,\gamma_j\rr =0 \quad \hbox{and} \quad
\omega_M(\gamma_i,\gamma_j) \equiv
\ll M,[\gamma_i,\gamma_j]\rr =0,\quad \forall
\gamma_i,\gamma_j \in \Gamma.
\eqno(5)
$$
When the constraints are first class, they generate gauge
symmetries on the constraint surface (2) through
the KM Poisson bracket.
For this reason, the subalgebra $\Gamma$ will be called a
{\it gauge algebra}.
It is natural to look at the gauge invariant functions,
namely those functions (weakly) commuting with the constraints.
Under certain technical conditions,
which include the non-degeneracy condition $[M,\gamma] \neq 0$ for
any
$\gamma \in \Gamma$, it has been shown in [3] that the set of
gauge invariant functions of the constrained current
has a basis which is
differential polynomial in the current components
and consists of a Virasoro
density and primary fields. Thus, when these technical
conditions
are satisfied, the gauge invariant functions form
a Zamolodchikov W-algebra under the KM Poisson bracket (1).

In the above
context the choice of constraints made by Bershadsky
for $sl(n,R)$ may be described as follows. Let $e_{r,s}$
denote the usual
one-entry generators of $gl(n,R)$ and $\D=\{e_{r,s}\}_{r<s}$
the upper triangular, maximal nilpotent subalgebra
of $sl(n,R)$. Then the constraints read ($1\leq l \leq n-1$)
$$
J_\D(x)= M+j_\D(x)\ ,
\qquad {\rm with}\qquad j_\D(x)\in \D^\perp\ ,
\eqno(6.{\rm a})
$$
where
$$
M= e_{l+1,1}+ e_{l+2,2} + \cdots + e_{n,n-l}.
\eqno(6.{\rm b})
$$
That is, the entries of the matrix $M$ are all zero
except those on a line parallel to the diagonal and $l$ steps
below it,
which are unity, and $J(x)$ is constrained to be upper triangular
apart
from a strictly lower triangular constant piece equal to $ M$.

The constraints (6) are not preserved by the standard Virasoro
density $L_{\rm KM}(x)$, but the modified Virasoro density
$L_H(x)$ in (4)
does preserve them
with $H$ given by the diagonal matrix
$H_l ={1 \over {2l}} \sum_{i=1}^n (n+1-2i) e_{ii}$.
In other words, $H_l$ gives all the elements on the $k$-th slanted
line above
(or below) the diagonal a grade $+{k \over l}$ (or $-{k \over l}$)
and, hence,
gives $M$ a grade $-1$.  This implies (3), that is,
$L_{H_l}(x)$ weakly commutes with all the constraints in (6),
and thus
whatever the reduction process is, the final reduced system is
going to be conformally invariant.

The case $l=1$ is the usual Toda case [1],
but for $l \geq 2$ there are two features not encountered in
previous reductions, as we shall see shortly.
First, the constraints are
not all first class because, although they satisfy the first
condition in (5), they violate the second one.
Second, the operator ${\rm ad}_M=[M ,\,\, \cdot \, ]$ has a
non-trivial kernel in $\D$.
Accordingly, the differential polynomial
gauge fixing algorithm developed in [3] for analyzing
reductions by first class, non-degenerate constraints cannot
be applied to the present situation.
On the other hand,
since $M$ is nilpotent, the $sl(2,R)$ structure is
intact, i.e., there should exist a set of $sl(2,R)$ generators
in which $M$ is identified with $M_-$.
Parametrizing $n=ml+r$ with $m=\bigl[{n \over l}\bigr]$
and $0 \leq r <l$,
a convenient choice of the other two generators is
$$
M_0 = {\rm diag}\Bigl(
      \overbrace{{m\over2}, \cdots}^{r \rm\;times},\,
      \overbrace{{{m-1}\over2}, \cdots}^{(l-r) \rm\;times},\,
      \cdots,
      \overbrace{-{m\over2}, \cdots}^{r \rm\;times} \Bigr),
\eqno(7)
$$
(the multiplicities, $r$ and $l-r$, occur alternately and end
with $r$) and
$$
M_+ = a_1\,e_{1,l+1} + a_2\,e_{2,l+2} + \ldots +
a_{n-l}\,e_{n-l,l},
\eqno(8)
$$
where the coefficients $a_i$ in (8) are given by
the first $n-l$ terms in the
following series ($k \geq 0$)
$$
\eqalign{
& a_{kl+1} = a_{kl+2} = \cdots = a_{kl+r} = (k+1)(m-k), \cr
& a_{kl+r+1} = a_{kl+r+2} = \cdots = a_{(k+1)l} = (k+1)(m-k-1).
\cr
}
\eqno(9)
$$
The meaning of (7) is that the fundamental of $sl(n,R)$
branches into $l$ irreducible $sl(2,R)$
representations (irreps.), $r$ of spin
${m \over 2}$ and $(l-r)$ of spin ${{m-1} \over 2}$.
{}From this, we get
that the adjoint of $sl(n,R)$ contains $(l^2m + r^2 -1)$
$sl(2,R)$ irreps.

{}From (7), we see that all the generators of $\D$ have an
$\ad_{M_0}$-eigenvalue
greater than or equal to zero. Thus the only
elements of $\D$ in the kernel of $\ad_{M_-}$ are necessarily
$sl(2,R)$ scalars. Let us denote this part of $\D$ by $\S$:
$$ \S = \{\s \in \D\;:\;[M_-,\s]=0\,,\,[M_0,\s]=0\}.
\eqno(10) $$
Using the expressions (6) and (7) of $M_-$ and $M_0$, we find
that, as
a concrete $n \times n$ matrix, any element $\s$ of $\S$ takes
the following block-diagonal form
$$\sigma=\hbox{block-diag} \{ \Sigma_0,\sigma_0,\Sigma_0,
\ldots ,\Sigma_0,\sigma_0,\Sigma_0 \},
\eqno(11)
$$
where $\Sigma_0$ and $\sigma_0$ are strictly upper triangular
$r \times r$ and $(l-r) \times (l-r)$
matrices, respectively,
which repeat themselves alternately.
{}From this it easy to compute that
$$
{\rm dim}\,\S = {1 \over 4}[l(l-2)+(l-2r)^2],
\eqno(12)
$$
which shows that for $l \geq 2$ the set
$\S$ is non-empty, except in the special case $l=2$, $r=1$,
that is,
$W^2_n$ with $n$ odd.

{}From the previous remarks, we can decompose the
subalgebra $\D$
according to the grading provided by the
eigenvalues of ${\rm ad}_{M_0}$:
$$
\D = \Delta_0 + {\cal G}_\half + {\cal G}_{\geq 1},
\qquad
{\cal G}_{\geq 1} = \sum_{i=1}^m {\cal G}_i,
\eqno(13)
$$
where $\Delta_0$ is the grade 0 subspace in $\D$.
This equation clearly shows that $\omega_{M_-}(\D,\D)=0$ is not
satisfied, which means that the constraints (6) defined by $\D$
are not first class.
Therefore we should separate this system of constraints into
first class and second class parts.
In fact, it follows from (1) that the first class part,
generating gauge transformations on the constraint surface (6),
is the one associated to the maximal subalgebra
${\cal D}\subset \Delta$
subject to $\omega_{M_-}({\cal D}, \Delta )=0$.
Explicitly, we obtain from (13) that
$$
{\cal D}=\S + {\cal D}_1 + {\cal G}_{>1},
\eqno(14)
$$
where ${\cal D}_1$ is that subspace of ${\cal G}_1$ for which
$\omega_{M_-}({\cal D}_1, \Delta_0)=0$.
The second class part then belongs to the
complementary space ${\cal C}$ entering into the
decomposition
$\Delta ={\cal D}+{\cal C}$, since the restriction of
$\omega_{M_-}$ to ${\cal C}$
is non-degenerate.
By combining (13) and (14), we see
that ${\cal C}$ naturally decomposes into
$$
{\cal C} = {\cal C}_0 + {\cal G}_{\half}+ {\cal C}_1 ,
\eqno(15)$$
where $\Delta_0 ={\cal D}_0+{\cal C}_0$ and
${\cal G}_1={\cal D}_1+{\cal C}_1$.

Although  ${\cal D}\subset \Delta$ defines the maximal set of
first class constraints which weakly commute with all
constraints belonging to $\Delta$, it may be enlarged to a bigger
subspace $\G$, which still defines a set of first
class constraints, by discarding troublesome elements in
$\D$ which do not comply with the condition (5).
This procedure has already been used in [3],
where it was called \lq the method of symplectic halving',
since the general idea is to find a gauge algebra of first
class constraints in the form $\Gamma ={\cal D}+{\cal P}$,
where ${\cal P}$ is defined in terms of an appropriate
direct sum decomposition (\lq halving') of
the second class part ${\cal C}$ into
symplectically conjugate subspaces,
${\cal C} = {\cal P}+{\cal Q}$.

To apply the above method, we note that, for $M_0$-grading reasons,
the 2-form $\omega_{M_-}$ is actually separately non-degenerate on
the two subspaces
${\cal G}_\half$ and
${\cal C}_0 + {\cal C}_1$
of ${\cal C}$ (15).
Thus we can take the ${\cal Q}$ and ${\cal P}$ subspaces
of ${\cal C}$ to be
${\cal Q} = {\cal Q}_{0} + {\cal Q}_{\half}$
and
${\cal P} = {\cal P}_{\half}+{\cal P}_1$
where, by definition,
${\cal Q}_0={\cal C}_0$, ${\cal P}_1={\cal C}_1$ and
${\cal G}_\half= {\cal P}_\half + {\cal Q}_\half$.
What is meant by this decomposition is that
$\omega_{M_-}$  vanishes
identically on each of the four subspaces
${\cal Q}_\half$, ${\cal P}_\half$, ${\cal Q}_0$ and
${\cal P}_1$, and that ${\cal Q}_\half$ and ${\cal P}_\half$ are
the $\omega_{M_-}$-duals of each other, as are ${\cal Q}_0$ and
${\cal P}_1$.
There is obviously a large freedom in choosing the symplectic
halving of ${\cal G}_{\half}$,
and it should be noted that at this stage we have not yet made
a specific choice.
Using these decompositions, we define
$\G \subset \D$ as
$$
\G = {\cal D} + {\cal P}_\half + {\cal P}_1 =
\S + {\cal P}_\half + {\cal G}_{\geq 1} ,
\eqno(16)
$$
and we wish to identify this $\Gamma$ as a gauge algebra of
first class constraints, the use of which is soon to be explained.

It is clear from the definition (16) that $\G$ satisfies the two
conditions in (5), however,
this definition does not automatically make
$\G$ a Lie algebra, since, for this to be the case, we must have
$$
[\S,{\cal P}_\half] \subset {\cal P}_\half.
\eqno(17)
$$
We now show that there exists (at least)
one halving of ${\cal G}_\half$ such that
$\G$ is indeed a Lie algebra.
For this, we introduce another grading
operator $\tilde H$ defined as follows.
As a diagonal matrix of
$sl(n,R)$, $\tilde H$ is obtained out of $M_0$ by first adding
$\half$ to its
half-integral eigenvalues, and then substracting a multiple
of the unit matrix so as to make the result traceless. Hence in
the fundamental of $sl(n,R)$, $\tilde H$ is given by
$$
\tilde H = M_0 - \lambda I_n + \cases{0 & on tensor irreps, \cr
                                  \half & on spinor irreps. \cr }
\eqno(18)
$$
The definition (18) makes clear that $\tilde H$ is an integral
grading,
commuting with $M_0$
and such that $[\tilde H,M_\pm]=\pm M_\pm$.  In the
adjoint of $sl(n,R)$, we then have $\ad_{\tilde H} =\ad_{M_0}$
on tensors, and
$\ad_{\tilde H}=\ad_{M_0} \pm \half$ on spinors.
In the last case,
$\ad_{\tilde H} -\ad_{M_0}$
equals $+\half$ as many times as $-\half$. In particular,
exactly half of ${\cal G}_\half$ has an $\tilde H$-grade
equal to 0, the
$\tilde H$-grade of the other half being 1. We then choose
the following
symplectic halving of ${\cal G}_\half$:
$$
{\cal Q}_\half
\equiv {\cal G}_\half \cap {\cal G}^{\tilde H}_0\,,
\qquad
{\rm and} \qquad
{\cal P}_\half
\equiv {\cal G}_\half \cap {\cal G}^{\tilde H}_1\, ,
\eqno(19)
$$
where the superscript $\tilde H$ means that the grades are
with respect to
$\tilde H$.
Since the
elements of $\S$, being tensor states,
have an $\tilde H$-grade equal to 0
(same value as their $M_0$-grade),
this choice of ${\cal P}_\half$ clearly
guarantees (17), that is, $\G$ is a Lie algebra.

Hence the subalgebra
$\G\subset\Delta$ given by (16) and (19) satisfies
all the conditions for defining first class constraints by means
of eq.(2).
Furthermore, it is easily seen that
the constraints belonging to
${\cal Q}$,
which are originally in $\D$ but are missing from $\G$, can be
recovered by regarding them as (partial) gauge fixing conditions
associated with
the piece ${\cal P}$ of $\Gamma$.
{}From this observation one concludes that
the reduced phase space,
obtained by imposing the first class constraints (2) and
quotienting by the $\Gamma$-gauge-transformations,
is identical with the reduced phase space
obtained by imposing Bershadsky's constraints (6) and
quotienting by the ${\cal D}$-gauge-transformations:
$$
\eqalignno{
\hbox{reduced phase space}
                           &= \{J_{\G}= M_- + j_{\G}(x)\}\,/\,
\{\G \hbox{-KM transformations}\} \cr
                           &= \{J_{\D}= M_- + j_{\D}(x)\}\,/\,
\{{\cal D}\hbox{-KM transformations}\}.
&(20)
}
$$
Correspondingly, the following elementary counting gives
the dimension of the $W_n^l$-algebras, i.e.,
the number of the gauge invariant degrees of freedom:
$$
\eqalign{
{\rm dim}\,W_n^l
&= {\rm dim}\,{\cal G} - \,{\rm dim}\,\G
                       - \,{\rm dim}\,\G
                    = {\rm dim}\,{\cal G} - {\rm dim}\,\D -
                                        {\rm dim}\,{\cal D} \cr
& = l(n + r + 1) - (l^2 + r^2 + 1). \cr
}
\eqno(21)
$$

\topinsert
\def\mrs#1{\smash{
                 \mathop{\longrightarrow}\limits^{#1}
                }
         }
\def\mrd#1{\smash{
                 \mathop{\Longrightarrow}\limits^{#1}
                }
         }

\def\mdd#1{\Big\Downarrow\
              \rlap{$\vcenter{\hbox{$\scriptstyle#1$}}$}
         }

\def\mud#1{\Big\Uparrow\
              \rlap{$\vcenter{\hbox{$\scriptstyle#1$}}$}
         }
$$
\matrix{
 & & & & J & \mrs{\Delta} & J_{\Delta}
       & \mrd{{\cal D}_{\rm gf}} & J^{\rm red}
       & \hbox{(Bershadsky)}
\cr \noalign{\medskip}
 & & & & & & \mud{{\cal P}_{\rm gf} = {\cal Q}}
 & & &
\cr \noalign{\medskip}
 & & & & J & \mrs{\Gamma} & J_{\Gamma}
                  & \mrd{\Gamma_{\rm gf}} & J^{\rm red} &
\cr \noalign{\medskip}
 & & & & & & \mdd{\tilde \Gamma_{\rm gf}} & & &
\cr \noalign{\medskip}
 J & \mrs{\tilde \Gamma}
   & J_{\tilde \Gamma}
   & \mrd{\tilde \Gamma_{\rm gf}}
   & J^{\rm hw}
   & \mrs{{\cal D}_0}
   & J^{\rm hw}_{{\cal D}_0}
   & \mrd{({\cal D}_0)_{\rm gf}}
   & J^{\rm red}
   & \hbox{(We)}
\cr
}
$$
\vskip 2mm
\baselineskip=10pt
\leftskip=1cm
\rightskip=1cm
\noindent
{\ninebf Figure 1.}
{\ninerm
Various processes reaching the same reduced phase space.
Single-lined arrows represent the imposition of constraints,
and double-lined arrows represent gauge fixings.
}
\vskip 5mm
\leftskip=0cm
\rightskip=0cm
\baselineskip=16pt
\endinsert


As usual, the reduced
phase space may be regarded
as the space of gauge-invariant functions of the constrained
currents and
a dim$W^l_n$-dimensional basis for the gauge-invariants
may be obtained by gauge-fixing. However,
the space $\G$ of the first class
constraints (16) contains a degenerate part ${\cal D}_0$,
which means
that it is impossible to perform the gauge fixing in the
previous manner [3].
Thus it is natural to first consider the non-degenerate
part of $\G$,
$$
\tilde\G = {\cal P}_\half + {\cal G}_{\geq 1},
\eqno(22)
$$
by the
usual procedure of gauge fixing,
and then find a convenient way of
gauge fixing for the degenerate part.
In doing this,
we are actually taking yet another path
leading to the reduced phase space, the
lower path in Fig.1.
This is again allowed because,
as we shall see below, the space of
current components eliminated by the
gauge fixing
for the non-degenerate part $\tilde\G$ can be chosen to be
independent of the degenerate part ${\cal D}_0$.

Now, imposing the constraints belonging to the
non-degenerate part $\tilde\G$,
$$
J_{\tilde\G}(x)= M_- + j_{\tilde\G}(x)\ ,
\qquad {\rm with}\qquad j_{\tilde\G}(x) \in \tilde\Gamma^\perp\ ,
\eqno(23)
$$
one sees from (22) that the space in which the current lies
can be explicitly given by
$\tilde\Gamma^\perp ={\cal G}_{\geq 0}+{\cal Q}_{-\half}$
with
${\cal Q}_{-\half}=[M_-,{\cal P}_\half ]$
being the subspace of ${\cal G}_{-{\half}}$ orthogonal to
${\cal P}_\half$.
One then finds that $\tilde\G^\perp$
can also be written in the form
$$
\tilde \Gamma^\perp =
            [M_-, \tilde\Gamma]+ {\rm Ker}({\rm ad}_{M_+})\ .
\eqno(24)
$$
The essential ingredient in fixing
the current in the usual form (called
\lq Drinfeld-Sokolov gauge' in [3]) is the following.
First, let us observe the gauge transformation,
$$
\eqalignno{
j_{\tilde\G}(x)
&\longrightarrow
e^{a(x)}\bigl( j_{\tilde\G}(x) + M_- \bigr) e^{-a(x)}
+ ( e^{a(x)} )^\prime e^{-a(x)} - M_- \cr
&\quad=
j_{\tilde\G}(x)  +  [a(x) , \, M_-]
+  [a(x) , \, j_{\tilde\G}(x)] + a'(x) + \cdots,
&(25)
}
$$
where $a(x) \in \tilde\Gamma$ is a local parameter which may be
decomposed into its grades, $a(x) = \sum_{i={1\over2}}^m a_i(x)$.
Concentrating in particular on the lowest grade component of the
gauge transformed current,
i.e., on the component in the space ${\cal Q}_{-\half}$,
we find that
the only contribution to the transformation
comes from the term
$[a_{1\over2}(x), \, M_-]$.  It then follows that as far as
this grade is concerned we can put the corresponding
current component to zero by a suitable choice of
the gauge
parameter $a_{1\over2}(x)$.
By using the non-degeneracy,
${\rm Ker(ad}_{M_-}) \cap \tilde\G = \{ 0 \}$,
and the graded structure (22),
it is also easy to see that,
by going up to spaces of higher
grades iteratively,
all the current components belonging to the space
$[M_-, \tilde\G]$ can be set to zero by choosing the $a_i(x)$'s
appropriately.
As a result, the current components which survive this
gauge fixing lie in a complementary space to
$[M_-,\tilde\G]$, which, in view of the decomposition (24),
we can take to be
Ker(${\rm ad}_{M_+}$), the space of the $sl(2,R)$
highest weight states in the adjoint of $sl(n,R)$.
Clearly, the gauge parameter
determined in the above procedure is
a differential polynomial
in the original current $j_{\tilde\G}(x)$.  This in turn implies
that the components $j^{\rm hw}(x)$
of the gauge transformed, highest weight, current
are also (gauge invariant) differential polynomials
$j^{\rm hw}(j_{\tilde \Gamma})$ when expressed in the
original current components.

As we have noted earlier,
the conformal invariance of our system is
guaranteed by choosing $H_l$ for the $H$ to define
the modified Virasoro
density (4).
However,
such an
$H$ is by no means unique,
as we have already encountered another
example $\tilde H$ in (18).
A natural choice
in dealing with the $sl(2,R)$ embedding is to take
$M_0$ itself to be the $H$ (which evidently satisfies (3)) and
use $L_{M_0}(x)$ to specify the conformal
structure of the theory.
The virtue of this choice lies in the fact that,
under the conformal transformation generated by $L_{M_0}(x)$,
all the surviving current components, $j^{\rm hw}(x)$,
except the Virasoro component turn out to be primary fields.
One can see this by explicitly computing the response
of the current components under the infinitesimal
conformal transformation
$\delta x = - f(x)$,
$$
\delta \, j^{\rm hw}(x)=f(x) (j^{\rm hw}(x))'+
f^{\prime}(x)\bigl( j^{\rm hw}(x)+[M_0, \,j^{\rm hw}(x)] \bigr)
-{1\over 2}f^{\prime\prime\prime}(x)M_+,
\eqno(26)
$$
which shows that the spin $s$ component of the highest weight
current has conformal weight $s+1$, except for
the $M_+$-component.  This component
turns out to be, up to $sl(2,R)$
scalars, $L_{M_0}(x)$ itself once reduced to the highest weight
gauge.

Since for every $\tilde\Gamma$-gauge-invariant
polynomial $P(j_{\tilde\Gamma})$ the invariance
implies
$P(j_{\tilde\Gamma})=P(j^{\rm hw}(j_{\tilde\Gamma}))$
it is clear that the
$j^{\rm hw}(j_{\tilde\Gamma})$
constitute a basis for the $\tilde \Gamma$-invariant polynomials.
It follows that the Poisson bracket algebra of the
$j^{\rm hw}(j_{\tilde\Gamma})$
closes
and since the
$j^{\rm hw}(j_{\tilde\Gamma})$
are primary and include a Virasoro density,
this is a
Zamolodchikov algebra. In fact it is the Zamolodchikov
algebra
considered for $W^2_4$ and $W^3_4$ in [8,9] and is a special case of
the ${\cal W}^{\cal G}_{\cal S}$ algebras
considered in [3].

The original constraints proposed by Bershadsky include,
however, the constraints corresponding to the subalgebra ${\cal
D}_0$ and we now have to consider the situation when these
constraints are imposed. It is not difficult to see
that when these constraints are imposed the current
$j^{\rm hw}_{{\cal D}_0}(x)$ takes the form
$$
j^{\rm hw}_{{\cal D}_0}(x)
= \pmatrix{
K_0 & K_{{\half}} & K_1        & K_{3 \over 2}&K_2    & ...&
K_{m-{\half}} & K_{m} \cr  \noalign{\medskip}
0   & k_0         & k_{{\half}} & k_1 & k_{3 \over 2} & ...&
k_{m-1} & k_{m-\half} \cr  \noalign{\medskip}
0   &  0          & K_0 & \star     & \star           & ...&
   \star     & \star     \cr   \noalign{\medskip}
0   &  0          & 0   & k_0   & \star               & ...&
  \star      & \star     \cr   \noalign{\medskip}
 ...&    ...      &  ...& ...&     ...                & ...&
   ...     &...  \cr   \noalign{\medskip}
0   &    0        &  0  &  0    &  0                  & ...&
    k_0    &  \star  \cr   \noalign{\medskip}
0   &    0        &  0  &  0    &   0                 & ...&
   0      & K_0   \cr
}
\eqno(27)
$$
where the
subscripts denote the spins with respect to $sl(2,R)$, the
$r \times r$ and $(l-r) \times (l-r)$  scalar matrices
$K_0(x)$ and $k_0(x)$
are upper triangular (including the diagonal)
and are identical copies, and the blocks
denoted with stars $\star$ are proportional to
the block-matrices $K_i(x)$
and $k_i(x)$ for $i\not= 0$, due to the highest weight condition
$[M_+,j^{\rm hw}(x)]=0$. The matrix (27) being in $sl(n,R)$,
the blocks
$K_0$ and $k_0$ are subjected to the condition $(m+1)\,
{\rm Tr}\,K_0
+ m \, {\rm Tr}\,k_0 = 0 $. All the entries in (27) are,
of course, gauge invariant with respect to the non-degenerate
(non-scalar) part $\tilde \Gamma$ of the gauge algebra. Note that
the Virasoro operator $L_{M_0}(x)$ is now linear in the $M_+$-component
and
quadratic in the diagonal scalar entries of $j^{\rm hw}_{{\cal
D}_0}(x)$. In (27) all the $\D$ constraints have been recovered,
some of them
as first class constraints, the others as $\tilde \G$-gauge-fixing
conditions.

The problem now is to find a basis for the functions of the currents
(27)
which are gauge-invariant with respect to the residual gauge-algebra
${\cal D}_0$, i.e. with respect to the gauge-transformations
$j^{\rm hw}_{{\cal D}_0}\rightarrow
S(x)(j^{\rm hw}_{{\cal D}_0}-\partial)S^{-1}(x)$
where $S(x)$ is the gauge-group generated by ${\cal D}_0$.
Note that, because the elements of ${\cal D}_0$
are scalars, the ${\cal D}_0$ gauge-transformations leave invariant
both the set
of highest weights and the set of non-highest weights and thus
preserve the form (27).
Furthermore they do not mix highest weights of different grades.

To find a basis for the set of gauge-invariant functions we follow the
usual procedure of gauge-fixing. The problem is that,
because $M_-$ does not appear in the above ${\cal D}_0$
gauge-transformation,
the gauge-parameters $\alpha$ cannot be determined as  polynomial
functions of the current.  The best one can do is to note that
that since
${\cal D}_0$ is nilpotent there are current components that transform
{\it linearly} in the $\alpha$'s, and use these to obtain the
$\alpha$'s as
simple fractions of current components. We shall call such gauges
{\it fractional} gauges. It is then clear that in a fractional gauge
the gauge-fixed components of the current will be gauge-invariant
{\it rational} functions of the current components in (27).

To see this explicitly we write the gauge group
elements as $ S(x) = e^{\s(x)} =
\hbox{block-diag}\{S_0,s_0,S_0, \ldots,s_0,S_0\}$, with
$S_0(x) =e^{\SS_0(x)}$
and $s_0(x) = e^{\s_0(x)}$
with respect to the block decomposition (11).
Then the
transformations of the different blocks $K_i$ and $k_i$ in
$j^{\rm hw}_{\S}(x)$ are
$$
\left\{\eqalign{& K_n \longrightarrow S_0 K_n S_0^{-1} +
                    S_0' S_0^{-1} \delta_{n,0}, \cr
                  & K_{n+\half} \longrightarrow S_0 K_{n+\half}
                    s_0^{-1}, \cr } \right.  \quad {\rm and}
\quad \left
\{\eqalign{& k_n \longrightarrow s_0 k_n s_0^{-1}+s_0's_0^{-1}
                    \delta_{n,0}, \cr
                  & k_{n+\half} \longrightarrow s_0 k_{n+\half}
                    S_0^{-1}. \cr } \right.
\eqno(28)
$$

We first consider the fixing of the parameters contained in $S_0$
which is an $r \times r$ upper triangular matrix. It can be written
as
$$
S_0 = S_0^{(2)} \cdot S_0^{(3)} \cdots S_0^{(r)}, \quad
{\rm with} \quad
S_0^{(i)} =
\pmatrix{
1    & 0   & ... &  0  & a_{1,i}   &  0  & ... & 0   \cr
0    & 1   & ... &  0  & a_{2,i}   &  0  & ... & 0   \cr
...  & ... & ... & ... & ...       & ... & ... & ... \cr
0    & 0   & ... &  1  & a_{i-1,i} &  0  & ... & 0   \cr
0    & 0   & ... &  0  & 1         &  0  & ... & 0   \cr
0    & 0   & ... &  0  & 0         &  1  & ... & 0   \cr
...  & ... & ... & ... & ...       & ... & ... & ... \cr
0    & 0   & ... &  0  & 0         &  0  & ... & 1   \cr
}. \eqno(29)
$$
The elements of $S_0^{(i)}$ form an $(i-1)$-dimensional
abelian subgroup of $S_0$ and is an invariant subgroup of
$S_0^{(2)} \cdot S_0^{(3)} \cdots S_0^{(i)}$ ($2 \leq i \leq r$).
In order to fix the gauge parameters $a_{i,j}(x)$, $i < j$,
contained in $S_0$, we consider the transformation of the
block $K_0$ and the first column of $K_\half$
(which always exists if $S_0$ does), which we parametrize as
$$
\pmatrix{K_0 \,\, \left| \,\, K_\half \right. \cr} =
\left(
\matrix{
*  &  * & * & ... &  *   &  \psi_1      & \chi_1     & \!\! \cr
0  &  * & * & ... &  *   &  \psi_2      & \chi_2     & \!\! \cr
0  &  0 & * & ... &  *   &  \psi_3      & \chi_3     & \!\! \cr
...& ...&...& ... & ...  &  ...         & ...        & \!\! \cr
0  &  0 & 0 & ... &  *   &  \psi_{r-2}  & \chi_{r-2} & \!\! \cr
0  &  0 & 0 & ... &  0   &  \psi_{r-1}  & \chi_{r-1} & \!\! \cr
0  &  0 & 0 & ... &  0   &    0         & \chi_{r}   & \!\! \cr
} \right|
\!\!
\left.
\matrix{
\quad \phi_1    & *   & *   & ... & *   \cr
\quad \phi_2    & *   & *   & ... & *   \cr
\quad \phi_3    & *   & *   & ... & *   \cr
\quad ...       & ... & ... & ... & ... \cr
\quad \phi_{r-2}& * & * & ... & * \cr
\quad \phi_{r-1}& * & * & ... & * \cr
\quad \phi_{r}  & * & * & ... & * \cr
} \right).
\eqno(30)
$$
We proceed recursively by examining
successively the first column of $K_\half$,
the last column of $K_0$, the second-last column of $K_0$, and so
on, thereby fixing the parameters of $S_0^{(r)}$, then those of
$S_0^{(r-1)}$, $S_0^{(r-2)}$, and so forth.
By the $S_0^{(r)}$ gauge transformation,
the first column of $K_\half$ becomes
$$
\phi_i \rightarrow \phi_i + a_{i,r}\phi_r\,,
\qquad
\phi_r \rightarrow \phi_r,
\eqno(31)
$$
for $1 \leq i \leq r-1$.
As a consequence, the choice
$$
a_{i,r}=-{\phi_i \over \phi_r}, \qquad 1 \leq i \leq r-1
\eqno(32)
$$
makes the components $\phi_i$, $1 \leq i \leq r-1$, vanish.
The other components of the current also undergo the $S_0^{(r)}$
transformations, and since the parameters (32) are
rational, so will be the components of the
$S_0^{(r)}$ gauge-fixed current. In general, they will contain a
dependence in $a_{i,r}$ and the first derivatives $a_{i,r}'$.
Although not polynomial, we note that these components
are still primary fields.
Indeed, from (32), all the parameters $a_{i,r}$ are
{\it conformal} scalars, because all the $\phi_i$'s are conformal
primary fields with the same conformal weight, namely,
${3 \over 2}$.
Therefore, as long as the non-derivative terms in
the transformations (28) are concerned, the entries of the $K$
and $k$
blocks retain their
primary field character. As to the derivative terms,
only the entries of
$K_0$, which are $sl(2,R)$ scalars and thus conformal
vectors, can possibly pick up a linear dependence in $a_{i,r}'$.
But
the parameters $a_{i,r}$ being conformal scalars, their first
derivatives
are conformal vectors, and so the entries of $K_0$ also
remain primary fields.
(In fact, a gauge transformation always preserves the conformal
weights if the charge of the transformation is conformally
invariant. In the present case this is equivalent to the condition
that the gauge parameter be a conformal scalar.)
So at this stage the components of the $S_0^{(r)}$
gauge-fixed current are all primary fields
(except the $M_+$-component) with
their conformal weight still
given by their $sl(2,R)$ grade plus one. However, they are {\it
rational} rather than polynomial, their
rational character coming from the denominator of the
parameters $a_{i,r}$, which is the gauge invariant
current component $\phi_r$.

We now go to the second step, and look at the
$S^{(r-1)}_0$ transformation of the $S^{(r)}_0$ gauge fixed
components of the last column of $K_0$.  These, of course, have
changed under the $S^{(r)}_0$ transformation, for
example, $\chi_{r-1}$ has become
$$
\tilde\chi_{r-1} =
\chi_{r-1} + \bigl( \psi_{r-1} - \chi_r \bigr) {\phi_{r-1}
\over \phi_r} - \bigl({\phi_{r-1} \over \phi_r}\bigr)'.
\eqno(33)
$$
For brevity we denote the (gauge-invariant) $S^{(r)}_0$ transformed
entries in the
column by $(\tilde\chi_1,\tilde\chi_2,\ldots,\tilde\chi_r)^t$.
The $S_0^{(r-1)}$ transformation yields
$$
\tilde\chi_i \longrightarrow
\tilde\chi_i + a_{i,r-1}\tilde\chi_{r-1}\,, \qquad
\tilde\chi_{r-1} \longrightarrow \tilde\chi_{r-1}\, ,\qquad
\tilde\chi_r \longrightarrow \tilde\chi_r\, ,
\eqno(34) $$
for $1 \leq i \leq r-2$.  One thus finds that
the first $(r-2)$ components can be gauged away by choosing
$$
a_{i,r-1}
= -{\tilde\chi_i \over
\tilde\chi_{r-1}}\,, \qquad 1 \leq i \leq r-2.
\eqno(35)
$$
As in the first step, the parameters $a_{i,r-1}$ given
by (35) are conformal scalars and the denominator
$\tilde\chi_{r-1}$ is a gauge invariant primary field. Thus
we obtain a basis of primary field rational functions which are
invariant under both $S_0^{(r)}$ and $S_0^{(r-1)}$ transformations.

We proceed in
the same way for the rest of the columns of $K_0$.
Namely, we use $S_0^{(r-2)}$ to remove
the first $(r-3)$ entries in the $(r-1)$-th column of $K_0$,
then use $S_0^{(r-3)}$
to remove the first $(r-4)$ entries in the
$(r-2)$-th column of $K_0$, and so on.
After the last step,
the fully constrained and $S_0$ gauge-fixed
current blocks $K_0$ and
$K_\half$ take the following form
$$
\pmatrix{
K_0 \,\, \left| \,\, K_\half \right. \cr} =
\left( \matrix{
\h   & \ss &  0   &  ... &  0  &   0  &   0  & \!\! \cr
0    & \h  & \ss  &  ... &  0  &   0  &   0  & \!\! \cr
0    &  0  & \h   &  ... &  0  &   0  &   0  & \!\! \cr
 ... & ... & ...  &  ... & ... &  ... &  ... & \!\! \cr
0    &  0  &  0   &  ... & \h  &  \ss &   0  & \!\! \cr
0    &  0  &  0   &  ... &  0  &  \h  & \ss  & \!\! \cr
0    &  0  &  0   &  ... &  0  &   0  &  \h  & \!\! \cr
} \right| \!\!\!
\left.
\matrix{
               \quad 0 & * & * & ... & * \cr
               \quad 0 & * & * & ... & * \cr
               \quad 0 & * & * & ... & * \cr
               \quad ...&...&... & ... &... \cr
               \quad 0 & * & * & ... & * \cr
               \quad 0 & * & * & ... & * \cr
               \quad \ss & * & * & ... & * \cr } \right).
\eqno(36)
$$
Here the symbols $\h$ in the diagonal line in $K_0$,
as well as in the lower-left corner of
$K_\half$ (which is $\phi_r$),
stand for the components which are $\S$ gauge invariant
from the beginning (see eq.(28)).
The rest of the components denoted by $*$ are still to be made
gauge invariant in the course of the gauge fixing of $s_0$
transformations. Exactly the same procedure can be used to fix
the parameters of $s_0$. In the first step, we use the first column
of
$k_\half$ (or $k_1$ if $r=0$ when $k_\half$ does not exist)
and proceed to the last column of $k_0$, and so on.
At the end of the procedure, when all the gauge parameters are fixed,
we are left with a set of $\dim W_n^l$ primary-field gauge-invariant
rational functions $j^{\rm red}(\j)$.
Now, for every $\Gamma$-invariant rational function
$R(j_{\Gamma})$ the gauge-invariance with respect to $\tilde \Gamma$
and ${\cal D}_0$ respectively imply that
$R(j_{\Gamma})=R(\j(j_{\gamma}))=R(j^{\rm red}(\j(j_{\gamma}))$,
 which shows that  the $j^{\rm red}$ constitute a
basis for the $\Gamma$-invariant rational functions. From this it
follows
that the Poisson bracket algebra of the $j^{\rm red}$ closes
rationally
and, since the $j^{\rm red}$ are primary, this algebra is a {\it
rational}
extension of the Virasoro algebra by primary fields.
Thus it is not quite a Zamolodchikov algebra in the
original sense of the word.

Although the basis is only rational one can improve the situation a
little by using the fact that
the denominators in the rational basis are separately
gauge-invariant and are primary fields. This
permits one to
convert the basis of rational functions into a basis
of polynomials without losing the
primary field character, as follows:
In the first step given by (32), some of the current components
(such
as $\tilde \chi_{r-1}$ in (33)) become rational functions with
either $\phi_r$ or $\phi_r^2$ in the denominator, so to convert them
to polynomials we simply multiply them by suitable powers of
$\phi_r$.
The field $\phi_r$ being itself a primary field, the multiplication
$\phi_r$ does not spoil the primary field property, but simply
shifts the values of the conformal weights. Thus, for example,
$\tilde \chi_{r-1}$ becomes
$$
\tilde\chi_{r-1} \, \phi_r^2 =
\chi_{r-1}\,
\phi_r^2 + \bigl( \psi_{r-1} - \chi_r \bigr) \phi_{r-1}
\, \phi_r - \phi_{r-1}'\,\phi_r + \phi_{r-1}\,\phi_r',
\eqno(37) $$
which is a gauge-invariant polynomial of conformal weight 4.
In the next step the denominators are either $\tilde \chi_{r-1}$ or
$\tilde \chi^2_{r-1}$ and we make the $j^{\rm red}$ polynomial by
multiplying
across by suitable powers of $\tilde \chi_{r-1}$. Again this field
is primary and does not spoil the primary field property of the
current components. The later steps are taken in
a similar manner.
The denominators $\phi_r$,
$\tilde \chi_{r-1}$ etc. used in the conversion are those marked by
the symbol $\ss$ on the slanted line just above the diagonal line
in the matrix (36).

The problem is that the basis of polynomials so constructed
is still only a rational basis, in the sense that there are
gauge-invariant
primary-field polynomials which cannot be expressed in terms of them
using only polynomial coefficients.
We illustrate this by considering the
$W^2_{2m}$ algebras, for which ${\cal D}_0$ is
one-dimensional and only the blocks $k_i$ for integer $i$
in the matrix (27) survive. If we write
$$k_0=x\sigma_3 +y\sigma_+ \quad \hbox{and} \quad
k_i=d_i + c_i\sigma_- +a_i\sigma_3 +b_i \sigma_+ , \qquad
1\leq i \leq m-1,  \eqno(38)$$
where the $\sigma$'s are the Pauli matrices, we see at once that
$\{x,c_i,d_i\}$ are ${\cal D}_0$-invariants
and the transformation properties of the remaining fields are
$$y \longrightarrow y-2\alpha x + \alpha' ,\qquad a_i \longrightarrow
a_i+\alpha c_i ,\qquad b_i \longrightarrow b_i - 2\alpha a_i -
\alpha^2 c_i,
\eqno(39) $$
where $\alpha$ is the gauge-parameter. Thus
if we gauge-fix by choosing $\alpha = -a/c$ for
the $\{a,c\}$ out of some particular block $k$, we obtain the
rational
basis
$$
x, \quad  y+2{ax \over c} -({a \over c})', \qquad c_i, \quad d_i,
\qquad a_i -{a \over c}c_i,  \qquad
b_i + 2{aa_i \over c} - ({a \over c})^2c_i.
\eqno(40)
$$
On conversion to polynomials these become $\{x,\, c_i, \, d_i\}$
and
$$
Y = c^2y+2acx +(c'a-a'c),\qquad
A_i = ca_i - ac_i, \qquad B_i = c^2b_i + 2aca_i - a^2c_i.
\eqno(41)
$$
But (41) is by no means a basis in which any gauge-invariant
polynomial can be
expanded using only polynomial coefficients since, for example,
$$
{\rm Tr}\,k_ik_j=2a_ia_j+b_ic_j+c_ib_j   \quad \hbox{and} \quad
{\rm Tr}\,\sigma_+k_ik_j = c_ia_j-c_ja_i
\eqno(42)
$$
for $k_i,k_j \neq k$ is a set of $(m-2)^2$ gauge-invariant
polynomials which, since they can also be written as
$$
{\rm Tr}\,k_ik_j= {1\over c^2}\bigl(2A_iA_j+B_ic_j+c_iB_j\bigr)
\quad \hbox{and} \quad
{\rm Tr}\,\sigma_+k_ik_j = {1\over c}\bigl(c_iA_j-c_jA_i\bigr),
\eqno(43)
$$
cannot be expanded
in the basis (41) using only polynomial coefficients.
Thus the Poisson-bracket
algebra of the polynomial basis (41) is not
guaranteed
to close polynomially.

Let us finally consider the Virasoro centre. The ambiguity of the
Virasoro density implies that the
conformal structure of the theory is not uniquely determined.
In particular, the
Virasoro centre appearing in the (quantum) $W_n^l$-algebras may
depend on the conformal structure, which we now examine.
For this, let us consider the BRST
system constructed from the constrained KM
theory in which a set of
ghosts are introduced associated
with the gauge algebra $\Gamma$ [3].
Then in the BRST system the Virasoro density
acquires the ghost part,
$L_{\rm tot}(x) = L_{M_0}(x) + L_{\rm ghost}(x)$,
and hence
the total Virasoro centre comprises the three contributions
$$
c_{\rm tot} = c_{\rm KM} + c_{\rm mod} + c_{\rm ghost},
\eqno(44)
$$
where $c_{\rm KM}$ is the usual centre coming from
$L_{\rm KM}(x)$,
and $c_{\rm mod}$ from the modification
in the Virasoro density.  If we adopt $L_{M_0}(x)$
in the present $W_n^l$ case, they read
$$
c_{\rm KM} = {{(n^2 - 1)k}\over{k+n}}, \quad
c_{\rm mod} = - 12k \ll M_0, M_0 \rr
        = - km(m+1)\bigl[3n - (2m+1)l \bigr],
\eqno(45)
$$
with $k$ being the KM level.
The ghost centre $c_{\rm ghost}$
arising from $L_{\rm ghost}(x)$ is
computed by using the standard formula,
$c(i) = -2 \bigl[ 1 + 6 i (i - 1) \bigr]$,
which gives the centre from
a pair of ghosts associated with a
grade $i$ element in $\Gamma$.
Taking into
account the multiplicities of the grades in $\Gamma$,
one finds
$$
\eqalignno{
c_{\rm ghost} &= c(0)\, {\rm dim\,}{\cal D}_0
          + c({1\over2})\,{\rm dim\,}{\cal P}_{1\over2}
          + \sum_{i=1}^{m}c(i)\,{\rm dim\,}{\cal G}_{i} \cr
        &= -(m^3 + 4m^2 + 3m + 1)l^2
          + \bigl[ n(2m^3 + 3m^2 + 6m + 2) + 1 \bigr] l \cr
        &\qquad - n^2 (3m^2 + 2).
&(46)
}
$$
As we have expected,
the result does not agree with the one obtained
by Bershadsky [6] who adopted
$L_{H_l}(x)$ in computing the Virasoro centre for $W_n^2$.

In this paper we have shown that the constraints introduced by
Bershadsky to define $W^l_n$-algebras are equivalent to a set of
first-class constraints with gauge-algebra of the form $\Gamma=
{\cal
D}_0 \wedge \tilde \Gamma$, where ${\cal D}_0$ is degenerate and
$\tilde \Gamma$ is non-degenerate. If ${\cal D}_0$ is zero (as
happens only for $W^2_n$ with $n$ odd) or is simply omitted, as
in [8,9], the natural basis for the gauge-invariant functions
of the constrained
currents is a set of polynomial primary fields
and these generate a Zamolodchikov algebra which is a special case
of the
${\cal W}^{\cal G}_{\cal S}$ algebras
discussed in [3]. If ${\cal D}_0$ is included, so as to
obtain the full set of Bershadsky constraints, then the natural
basis for the gauge-invariant functions is a set of
primary fields, which are rational functions of the constrained
currents, and we have given an algorithm for
their construction. The associated Poisson-bracket algebra is a
rational, rather than polynomial,  extension of the Virasoro algebra
by primary  fields, and is thus not quite a Zamolodchikov algebra.
There is a natural mechanism for replacing the rational primary-field
basis by a polynomial primary-field basis, but not every
gauge-invariant
polynomial can be polynomially-expanded in the new basis (i.e.
expanded using polynomial coefficients) and thus the associated
Poisson-bracket algebra is not guaranteed to close polynomially.
One may, of course, extend the polynomial basis to a (much larger)
one,
in terms of which the gauge-invariant polynomials can be
polynomially-expanded. But then the expansion is not unique, and
it is
an open question as to whether there exists any dim$W^l_n$ subset
whose Poisson-brackets close to form a Zamolodchikov algebra. This
question is technically
difficult
to answer because the current components $\j(j_{\tilde\Gamma})$
used for the
${\cal D}_0$ reduction satisfy a
$W$ algebra (a ${\cal W}^{\cal G}_{\cal S}$ algebra)
rather than a Kac-Moody algebra.

\eject
\centerline{\bf References}

\vskip 0.8truecm

\item{[1]}
J. Balog, L. Feh\'er, L. O'Raifeartaigh, P. Forg\'acs
and A. Wipf,
{\sl Ann. Phys.} (N. Y.)  {\bf 203} (1990) 76;
{\sl Phys. Lett.} {\bf 244B} (1990) 435.
\item{[2]}
F. A. Bais, T. Tjin and P. Van Driel, {\sl Nucl. Phys.} {\bf B357}
(1991) 632.
\item{[3]}
L. Feh\'er, L. O'Raifeartaigh, P. Ruelle, I. Tsutsui and A. Wipf,
{\it
On the General Structure of Hamiltonian Reductions of the WZNW
Theory},
Dublin preprint DIAS-STP-91-29.
\item{[4]}
A. N. Leznov and M. V. Saveliev,
{\sl Lett. Math. Phys.} {\bf 6} (1982) 505;
{\sl J. Sov. Math.} {\bf 36} (1987) 699;
\item{}
M. V. Savaliev, {\sl Mod. Phys. Lett.} {\bf A5} (1990) 2223.
\item{[5]}
A. Bilal and J.-L. Gervais, {\sl Phys. Lett.} {\bf 206B} (1988) 412;
{\sl Nucl. Phys.} {\bf B314} (1989) 646;
{\sl Nucl. Phys.} {\bf B318} (1989) 579.
\item{[6]}
M. Bershadsky, {\sl Commun. Math. Phys.} {\bf 139} (1991) 71.
\item{[7]}
A. M. Polyakov,
{\sl Int. J. Mod. Phys.} {\bf A5} (1990) 833.
\item{[8]}
I. Bakas and D. Depireux,
{\it Self-Duality, KdV flows and W-algebras}, Proceedings of the
{\it XX}-th International Conference on Differential Geometric
Methods in Theoretical Physics, New York, June 1991.
\item{[9]}
D. Depireux and P. Mathieu,
{\it On the Classical $W_n^l$-Algebras},
preprint LAVAL PHY-27/91.

\bye